\documentclass[sigconf]{acmart}

\AtBeginDocument{%
  \providecommand\BibTeX{{%
    \normalfont B\kern-0.5em{\scshape i\kern-0.25em b}\kern-0.8em\TeX}}}

\acmYear{2025}\copyrightyear{2025}
\setcopyright{acmlicensed}
\acmConference[FSE '25]{33rd ACM International Conference onq the Foundations of Software Engineering}{June 23--28, 2025}{Trondheim, Norway}
\acmBooktitle{33rd ACM International Conference on the Foundations of Software Engineering (FSE '25), June 23--28, 2025, Trondheim, Norway}
\acmDOI{XX.XXXX/XXXXXXX.XXXXXXX}
\acmISBN{XXX-X-XXXX-XXXX-X/XX/XX}

\acmBooktitle{Companion Proceedings of the 33rd ACM Symposium on the Foundations of Software Engineering (FSE '25), June 23--27, 2025, Trondheim, Norway}

\usepackage{multicol}
\usepackage{graphicx}
\graphicspath{ {./} }
\usepackage{array}
\usepackage{xcolor}
\usepackage{listings}
\definecolor{light-gray}{gray}{0.95}
\usepackage{multirow}
\usepackage{amsmath}
\usepackage{adjustbox}
\begin{document}

\title{Towards Adaptive Software Agents for Debugging}



\author {Yacine Majdoub}
\affiliation{
  \institution{University of Gabes, Tunisia}
  \city{}
  \country{}
}
\email{yacine.majdoub@enig.rnu.tn}

\author {Eya Ben Charrada}
\affiliation{
  \institution{University of Gabes, Tunisia}
  \city{}
  \country{}
}
\email{eya.bencharrada@fsg.rnu.tn}

\author {Haifa Touati}
\affiliation{
  \institution{University of Gabes, Tunisia}
  \city{}
  \country{}
}
\email{haifa.touati@univgb.tn}

\renewcommand{\shortauthors}{---}

\begin{abstract}
Using multiple agents was found to improve the debugging capabilities of Large Language Models. However, increasing the number of LLM-agents has several drawbacks such as increasing the running costs and rising the risk for the agents to lose focus. In this work, we propose an adaptive agentic design, where the number of agents and their roles are determined dynamically based on the characteristics of the task to be achieved.  In this design, the agents roles are not predefined, but are generated after analyzing the problem to be solved. Our initial evaluation shows that, with the adaptive design, the number of agents that are generated depends on the complexity of the buggy code. In fact, for simple code with mere syntax issues, the problem was usually fixed using one agent only. However, for more complex problems, we noticed the creation of a higher number of agents. Regarding the effectiveness of the fix, we noticed an average improvement of 11\% compared to the one-shot prompting. Given these promising results, we outline future research directions to improve our design for adaptive software agents that can autonomously plan and conduct their software goals. 

\end{abstract}




\maketitle

\section{Introduction}


Agentic AI systems, are \textit{"systems that can pursue complex goals with limited direct supervision"}~\cite{shavit2023practices}. Although there are various approaches that use agents to solve complex software problems~\cite{sakib2023extendingfrontierchatgptcode,wu2023largelanguagemodelsfault,lee2024unifieddebuggingapproachllmbased,hong2024metagptmetaprogrammingmultiagent,qin2024agentflscalingllmbasedfault,levin2024chatdbgaipowereddebuggingassistant,chen2023universalselfconsistencylargelanguage}, most of them lack in flexibility. In fact, existing approaches use static designs, where the number and roles of the agents are predefined in advance. 
This results in the agentic system being very limited in terms of adaptability to the  characteristics of the problem to be solved. 
For instance, static role design often lead to inefficient profiling, as agents may lack the specialization needed for specific tasks~\cite{han2024LLMmultiagentsystemschallenges}.
Additionally, static configuration, with a fixed number of agents, limits scalability and makes it difficult for systems to adapt to the diverse requirements of large-scale problems. This leads to unnecessary agent involvement in simple tasks, while causing bottlenecks in more complex scenarios~\cite{zhang2024cutcrapeconomicalcommunication}. Addressing these limitations through dynamic role assignment and scalable architectures is essential to improve resource efficiency and adaptability in agentic systems.

In this work, we propose a novel agentic design that combines three LLM reasoning techniques, which are planning, multi-agent collaboration, and iterative reflection, to develop a flexible agentic system for debugging, capable of dynamically adapting its strategy to the specific problem at hand. The proposed design includes two types of agents: (1) a central agent (main agent) that has the responsibility of managing the debugging process and (2) specialized agents that operate under the instructions of the main agent. The main agent performs an analysis of the buggy code, formulates an appropriate debugging strategy, and defines the profiles of specialized agents required to execute the task. These specialized agents are then created dynamically based on the profiles specified by the main agent. This adaptive architecture allows the system to adapt the used resources, in terms of number and roles of agents, to the complexity and requirements of the debugging task. 
Our initial evaluation demonstrates the ability of our proposed approach to dynamically adapt the debugging strategy and the number of used agents depending on the complexity and the requirements of each problem. Additionally, our approach showed significant improvement when applied across various LLMs, with an increase in bug fixing rate ranging from 6\% to 18\% depending on the model, outperforming traditional non-agentic debugging methods.

\section{background and motivation}
\label{sec:background}
The use of reasoning has been found to improve the debugging capabilities of LLMs; nevertheless, it poses several challenges. In this section, we first present  reasoning techniques for LLMs, then we review the use of these techniques in debugging, and finally we discuss their limitations and the challenges they pose. 

\subsection{Reasoning with LLMs}
\label{sec:reasoning}
Several techniques have been designed to improve the reasoning capabilities of LLMs. These approaches, which some call agentic design patterns~\cite{agenticDesignP}, aim to enhance the models' ability to handle complex tasks that require logical thinking, problem solving, and structured decision-making. In this section, we present some of the most popular techniques.

\paragraph{Chain-of-Thought} 
In this technique, the LLM is prompted to reason through tasks step by step, explicitly articulating intermediate reasoning rather than jumping directly to the final answer~\cite{wei2022chain}. This structured reasoning helps the model tackle complex problems by breaking them down into smaller, more manageable steps\cite{dutta2024thinkstepbystepmechanisticunderstanding}.

\paragraph{Self-consistency} 
Built on top of the Chain-of-Thought prompting~\cite{wang2022self}, this technique improves the quality of the output by generating multiple solutions for the same task and selecting the most consistent one. Instead of relying on a single response from an LLM, this approach involves prompting the LLM to produce several variations of the output, analyzing them, and then identifying the version that aligns most closely with a set of predefined  consensus~\cite{chen2023universalselfconsistencylargelanguage}.


\paragraph{Iterative-reflection}
Instead of producing a single final output, the LLM generates an initial response, evaluates it critically, and revises it based on constructive feedback~\cite{madaan2023selfrefineiterativerefinementselffeedback}. This process involves prompting the LLM to assess its output for issues such as correctness, clarity, and efficiency, and then using the feedback to refine the response~\cite{gou2024criticlargelanguagemodels}. The cycle can be repeated multiple times to achieve further improvements.

\paragraph{Planning}
The LLM autonomously determines a sequence of steps to complete a complex task, Huang et al.~\cite{huang2024understandingplanningLLMagents} classify LLM-based agent planning into five categories: (1) task decomposition, where the LLM decomposes the task into several sub-tasks, (2) multi-plan selection, where the LLM generates multiple alternative plans then selects one to execute, (3) external module-aided planning, where an external planer is used to evaluate the planning procedure, (4) reflection and refinement, where the LLM reflects on failures and refine the plan and (5) memory-augmented planning, where the LLM is augmented with an extra memory module that stores valuable information.

\paragraph{Tool use}
The LLMs can be  equipped with functions to gather information, perform actions, or manipulate data. This extension enables LLMs to carry out tasks like web searches or code execution to enhance their responses\cite{gao2025efficienttoolusechainofabstraction}. Examples of \textit{tool use}  include accessing external data sources (e.g., Stack Overflow) and using Retrieval-Augmented Generation (RAG) systems that dynamically searches for context-relevant information~\cite{patil2023gorillalargelanguagemodel}.

\paragraph{Multi-agent collaboration}
In this technique, different agents are created and work together to complete a complex task. Each agent focuses on its assigned sub-task, optimizing performance and making complex workflows more manageable~\cite{guo2024largelanguagemodelbased}. Although all agents may use the same LLM, they are prompted differently to simulate specialization, such as instructing one agent to prioritize efficient code fixes while another focuses on testing edge cases. This decomposition mirrors real-world team collaboration, where roles are assigned to specialists to handle different parts of a project~\cite{han2024LLMmultiagentsystemschallenges}.

\subsection{Use of LLM reasoning for debugging}
In this section, we review LLM-based research works on debugging. We summarize the reasoning techniques used by each of these works in Table~\ref{tab:approaches}.

Levin et al.~\cite{levin2024chatdbgaipowereddebuggingassistant} propose a debugger assistant named ChatDBG that uses the chain-of-thought technique by asking the assistant to explain the reasoning for the proposed fix. Chen et al.~\cite{chen2023teachinglargelanguagemodels} use "Self-Debugging", where LLMs autonomously refine the code based on execution outcomes. Qin et al.~\cite{qin2024agentflscalingllmbasedfault}
propose a mutli-agent system called AgentFL to localize bugs. AgentFL uses four agents with the roles Test Code Reviewer, Source Code Reviewer, Software Architect, and Software Test Engineer.
Ahmed et al.~\cite{ahmed2023better} use self-consistency to improve patching with LLMs.
Kang et al.~\cite{kang2023explainableautomateddebugginglarge} propose  AutoSD which uses reasoning to generate hypotheses about the source of the bug and then uses a debugger to verify the hypothesis.
Lee et al.~\cite{lee2024unifieddebuggingapproachllmbased} introduced FixAgent, a debugging framework, which combined multi-agent collaboration and tool use patterns to simulate real-life developers workflows. 
Similarly, MetaGPT, by Hong et al~.\cite{hong2024metagptmetaprogrammingmultiagent} encodes workflows as prompts, facilitating role-based collaboration among agents for tackling complex tasks.
The use of iterative refinement with chatGPT was explored by Sakib et al~\cite{sakib2023extendingfrontierchatgptcode} who found that it was very limited. 

\begin{table}[tb]
\caption{Comparison of used reasoning techniques across related approaches}
\label{tab:approaches}
\begin{adjustbox}{max width=0.48\textwidth,center}{
\renewcommand{\arraystretch}{1.5}%
\begin{tabular}{l|c|c|c|c|c|c}
\hline
Approach & \begin{tabular}[c]{@{}c@{}}Chain of \\ Thought\end{tabular} & \begin{tabular}[c]{@{}c@{}}Self \\ Consistency\end{tabular} & \multicolumn{1}{c|}{\begin{tabular}[c]{@{}c@{}}Iterative \\ Reflection\end{tabular}} & Planning & \begin{tabular}[c]{@{}c@{}}Tool \\ Use\end{tabular} & \begin{tabular}[c]{@{}c@{}}Multi\\ Agents\end{tabular} \\ \hline
Levin et al.~\cite{levin2024chatdbgaipowereddebuggingassistant} & \checkmark &  &  &  &  &  \\ \hline
Qin et al.~\cite{qin2024agentflscalingllmbasedfault} &  &  &  &  &  & \checkmark \\ \hline
Ahmed et al.~\cite{ahmed2023better} &  & \checkmark &  &  &  &  \\ \hline
Kang et al.~\cite{kang2023explainableautomateddebugginglarge} &  &  &  &  & \checkmark & \checkmark \\ \hline
Chen et .~\cite{chen2023teachinglargelanguagemodels} &  &  & \checkmark &  & \checkmark &  \\ \hline
Lee et al.~\cite{lee2024unifieddebuggingapproachllmbased} &  &  &  &  & \checkmark & \checkmark \\ \hline
Hong et al.~\cite{hong2024metagptmetaprogrammingmultiagent} & \checkmark &  &  &  &  & \checkmark \\ \hline
\end{tabular}
}\end{adjustbox}
\end{table}
 \subsection{Limitations of debugging with LLMs}
In this part, we first define the meaning of agenticness, then we discuss the agenticness of current debugging approaches and highlight the need for an adaptive agentic design.

 \paragraph{Meaning of agenticness}
Agenticness is the “degree to which a system can adaptably achieve complex goals in complex environments with limited direct supervision”~\cite{shavit2023practices}. Shavit et al.~\cite{shavit2023practices} break the agenticness property into four main components, namely: goal complexity, environmental complexity, adaptability, and independent execution.

\paragraph{Agenticness of current solutions}
When evaluating the performance of current LLM-based debugging techniques, we see that they are mostly focusing on the first component, i.e., goal complexity. In fact, most reviewed works aim at improving the abilities of LLMs to solve debugging tasks that are more and more complex using reasoning.
For the second component — environmental complexity — there were a couple of approaches that aim at gathering and using information from the environment such as the IDE tools. 
For the last two components  — adaptability and independent execution —  all current approaches have very limited capabilities. In fact, most current research  uses the same process for all types of bugs.  This makes the solutions extremely expensive to run and for solutions that include multi-agents, the performance might decrease when applied to simpler bugs. This is mainly due to the agents losing the focus on the main problem and being hard to predict.

\paragraph{Need for adaptive designs} Although recent LLM-based approaches allow solving more complex problems, they suffer from low adaptability to the problem characteristics. This makes them not only costly to run but also might reduce their effectiveness for certain types of problems. Therefore there is a need for a new agentic design that adapts the solution plan to the characteristics of the problem to be solved.

\section{Designing Adaptive agents}
\label{sec:adaptiveDesign}
\subsection{Overview}
\begin{figure}[tb]
\centering
\includegraphics[scale=0.31]{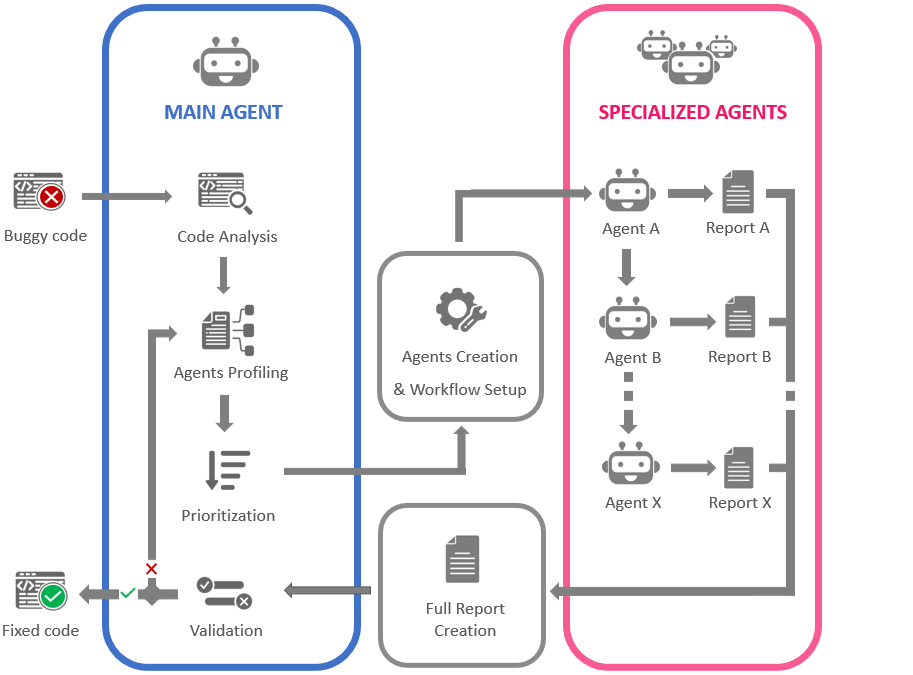}
\caption{Overview of the approach}
\label{fig:system}
\Description{Overview of the approach}
\end{figure}

To be adaptive, a system needs to be capable of adapting its approach to the problem's characteristics.  Therefore, the steps taken by the system should not be deterministic but should be decided dynamically based on the problem at hand. For this, we propose a design where we use the reasoning capabilities of LLMs to analyze the problem and plan the tasks and agent roles needed to address it. An overview of our design solution is presented in Figure~\ref{fig:system}. 

In this design, we use a combination of the   techniques \textit{planning, iterative reflection} and \textit{multi-agent collaboration}. For the planning, we use a \textit{decomposition-first method}~\cite{huang2024understandingplanningLLMagents}, where the task is first decomposed into sub-tasks then a plan is generated for each sub-task.  So a \textit{main agent} starts by defining the steps needed to solve the problem then it defines the requirements for \textit{specialized agents} that need to be created. Next, the specialized agents collaborate to carry out the tasks and  send a report to the main agent. The main agent   checks the outcome of the agents' work and decides whether to go for a new iteration. 
In the remainder of this section, we detail the functioning of the main and specialized agents. 

\subsection{Main agent / Team Leader}

The main agent plays the role of a Team leader. It is responsible for managing the flow of the task by dynamically guiding the creation and prioritization of the specialized agents, and the validation of the results. 
Rather than performing specific debugging tasks itself, the main agent makes the decisions about which agents to deploy and how they should be organized based on the type of the problem. 
The main agent is also responsible for integrating feedback from all agents, adjusting strategies, and guiding the system towards the efficient solution. In the next paragraphs, we detail the tasks of the main agent.

\paragraph{Code Analysis}
Code analysis is the initial step of the process in which the input code is analyzed to identify the type of problem that needs to be addressed. This process is critical because it guides the creation of the debugging strategy and the specification of the necessary agents. The main agent begins by scanning the code to identify common patterns or anomalies. For instance, unmatched parentheses or missing semicolons are identified as syntax errors, while issues like incorrect program behavior or unexpected outputs are flagged as logic bugs. Once the code is analyzed, the main agent outputs the type of issue along with a description of the problem. This analysis is essential because it is used to define the profiles of the agents needed to solve the problem.

\paragraph{Agent Profiling}
Based on the type of issues identified in the previous phase, the main agent creates the profiles of the specialized agents. For each profile, the main agent generates the prompt needed to configure the agent to perform the task at hand. For example, when a syntax error is detected, the main agent generates a prompt that will tell the specialized agent to have the role of a syntax checker and that it has to check the code, repair it, and generate a report about the fix.

\paragraph{Agent Prioritization}
After specifying the profiles of the agents, the main agent orders how the different agents will perform their tasks. The main-agent orders the specialized agents based on the relations between the tasks to be done. For example, if the analysis identifies a syntax error that prevents the code from running, a syntax checker is given higher priority over other agents, such as a semantic verifier, which may only be relevant after the code is syntactically correct.

\paragraph{Validation} After all specialized agents submit their reports, the main agent evaluates the outcomes of the executed tasks to determine whether the debugging process has successfully fixed all bugs. If the code is validated as corrected, the system outputs the corrected version to the user. However, if the issue persists, the main-agent goes back to the agent profiling and agent prioritization stages, in order to develop a new strategy to fix the issue. When defining the new strategy, the main agent takes into account the reports returned by the specialized agents To avoid having the system redoing the same plan in the new iteration,  we explicitly prompt the main agent to propose a debugging strategy that is different from the ones in the previous iterations.

\subsection{Specialized Agents / Team members }

The specialized agents are created based on the profiles that the main agent generated after analyzing the problem. Each agent has a specific role and well defined task, such as fixing syntax errors, detecting logical flaws, etc. 
Once created, the agent works  autonomously on the task assigned to it, using its specialized knowledge. Upon completing the task, the agent produces a report with its findings and recommendations that will be delivered later on to the main agent for validating the solution. The report also helps the main agent track the progress on the issue and generate a new plan if the solution is not validated.

\section{Preliminary evaluation}
\label{sec:evaluation}
To prove the feasibility  of our approach, we did a preliminary evaluation using 50 buggy code instances and four LLMs. We present the experiment design and the results in the next paragraphs.

\subsection{Experiment design}
\paragraph{Dataset} For the initial evaluation, we used a dataset consisting of 50 buggy code instances from the benchmark DebugBench~\cite{tian2024debugbench}. The code instances, which are written in python, are solution to coding problems from LeetCode\footnote{https://leetcode.com/}. Each instance includes different types of bugs with different levels of complexity. The bugs are categorized into four types of bugs: syntax errors, reference errors, logical errors and multiple errors . The complexity level of the instances ranges from simple problems requiring basic syntax fixes to more advanced issues that involve algorithmic logic and optimization challenges. 

\paragraph{Implementation} For the implementation, we wrote scripts that followed the adaptive design, where an LLM is prompted to be the main agent, which does the analysis of the code and creates the agents profiles. Then, based on the generated profiles, the script creates specialized agents, and orders the execution of the agents tasks according to the plan. Subsequently, the script prompts the main agent with the reports of the specialized agents and asks it to validate the solution or regenerate a different plan based on the output of the previous iteration. We run our script using four LLMs: Llama3, DeepSeek-coder, Mistral-Large and GPT-4, to compare their performance. The selection of the models was based on their demonstrated effectiveness in debugging~\cite{Majdoub_2024,tian2024debugbench,wu2023largelanguagemodelsfault}. 

\paragraph{Metrics} To evaluate the debugging performance of our solution, we counted the number of buggy instances that were fixed correctly. To evaluate the adaptability of our solution, we counted the number of agents created and iterations performed for each problem and compared it to the problem complexity.

\subsection{Results}

\begin{table}[tb]
\caption{Bug fix rate across different models.}
\label{tab:results}
\begin{adjustbox}{max width=0.48\textwidth,center}{
\renewcommand{\arraystretch}{1.2}%
\begin{tabular}{l|cc|c}
\hline
\multirow{2}{*}{Model}  & \multicolumn{2}{c|}{Bug Fix Rate}              & \multirow{2}{*}{Gain} \\ \cline{2-3}
                        & \multicolumn{1}{c|}{Base Model} & Adaptive Design &                       \\ \hline
Llama3                  & \multicolumn{1}{c|}{26/50}       & 35/50         & 18\%                  \\ \hline
DeepSeek-coder-Instruct & \multicolumn{1}{c|}{32/50}       & 38/50         & 12\%                  \\ \hline
Mistral-Large-Instruct  & \multicolumn{1}{c|}{29/50}       & 33/50         & 8\%                   \\ \hline
GPT-4                   & \multicolumn{1}{c|}{41/50}       & 44/50         & 6\%                  \\ \hline
\end{tabular}
}\end{adjustbox}
\end{table}

Our initial result suggest that the adaptive design has two advantages: (1) improved fix rate compared to the one-shot prompting and (2) an adapted usage of resources, in terms of the number of agents and iterations that were used.

\paragraph{Fix rate}  When comparing the fix rate of our solution to the one shot prompting, we noticed  an increase in the fix rate that ranged from 6\% to 18\% depending on the model. 
We report the detailed fix rates in Table~\ref{tab:results}. From results, we conclude that our adaptive design has the potential to improve the bug fixing performance of LLMs. 

\begin{figure}[tb]
\hspace*{-0.7cm}
\includegraphics[scale=0.27]{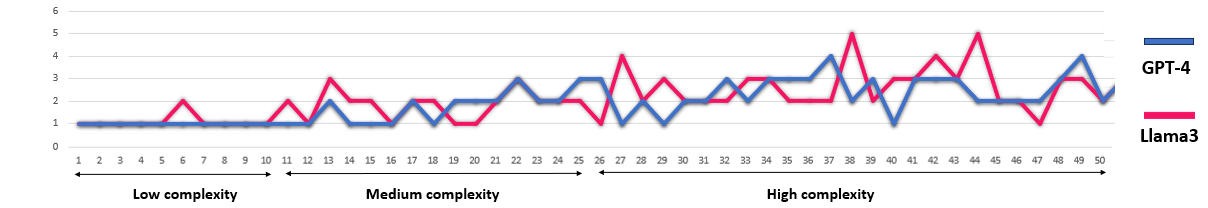}
\caption{Number of created specialized agents using Llama3 and GPT-4 for different complexity levels }
\label{fig:chart}
\Description{Number of created specialized agents}
\end{figure}

\begin{figure}[tb]
\hspace*{-0.7cm}
\includegraphics[scale=0.27]{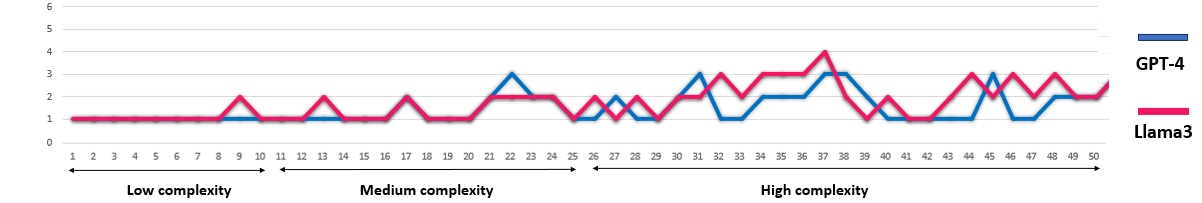}
\caption{Number of needed iterations using Llama3 and GPT-4 for different complexity levels }
\label{fig:chart2}
\Description{Number of iterations}
\end{figure}

\paragraph{Number of agents and iterations} The number of specialized agents that were created by our solution varied depending on the task. The number of created agents and needed iterations for each debugging task are shown in Figure \ref{fig:chart} and Figure \ref{fig:chart2} respectively . We see that for most low-complexity tasks, a single agent was sufficient. Medium-complexity tasks often required more collaboration, where the system created multiple agents, up to a maximum of three. High-complexity tasks often involved a higher number of agents and required multiple iterations, with a maximum of five agents engaged in some cases.

\section{Future Plans}
\label{sec:nextSteps}
Given the encouraging results of the initial evaluation, we plan to explore it further and adapt it to other software engineering tasks. In this section we summarise the main research directions that we plan to pursue.

\textit{Improving design choices.}   Although the current implementation has given promising results, we know that with LLM-based approaches, small changes can make big differences. In fact, any changing in the prompt might completely change the result. Therefore, we are planning to explore variations of our design and try different prompts to find out what implementation choices can give the best results in terms of solving the problem with adapted resource usage. We also plan to combine multiple LLMs, both open- and closed-source, and get the best possible results from their collaboration.

\textit{Extensive evaluation.}  We plan to extend the evaluation to a larger set of bugs. We will explore the effectiveness of different design choices on the correctness of the fix as well as the resources used for each type of problem. In addition to the quantitative evaluation, we will qualitatively analyze how the agents behave with different types of problems.

\textit{Adaptive Software Agent.}  We will explore the effectiveness of our adaptive agent design in performing other software engineering tasks, such as eliciting requirements, generating code or testing an implementation. Our ultimate goal is to have adaptive software agents that can adequately address any type of software engineering problems.  

\textit{Our source code is available at:} \url{https://github.com/YacineMajdoub/Adaptive-Agents-for-Debugging.git}

\newpage

\onecolumn \begin{multicols}{2}

\bibliographystyle{ACM-Reference-Format}

\bibliography{FSE}

\appendix

\end{multicols}
\end{document}